\documentclass[aip, twocolumn]{revtex4}%
\usepackage{amsfonts}
\usepackage{amsmath}
\usepackage{amssymb}
\usepackage{color}
\usepackage{graphicx}
\providecommand{\U}[1]{\protect\rule{.1in}{.1in}}

\begin{document}
\title{Prediction of extremely long mobile electron spin lifetimes at room temperature in wurtzite semiconductor quantum wells}
\author{N. J. Harmon}
\affiliation{Department of Physics, Ohio State University, 191 W. Woodruff Ave.,
Columbus, OH 43210, USA}
\author{W. O. Putikka}
\affiliation{Department of Physics, Ohio State University, 191 W. Woodruff Ave.,
Columbus, OH 43210, USA}
\author{Robert Joynt}
\affiliation{Department of Physics, Univeristy of Wisconsin-Madison, 1150 Univ. Ave.,
Madison, WI 53706, USA}

\begin{abstract}
Many proposed spintronics devices require mobile electrons at room temperature with long spin lifetimes. 
One route to achieving this is to use quantum wells with tunable spin-orbit (SO) parameters. 
Research has focused on zinc-blende materials such as GaAs which do not have long spin lifetimes at room temperature. We show that wurtzite (w) materials, which possess smaller SO coupling due to being low-Z, are better suited for spintronics applications.
This leads to predictions of spin lifetimes in w-AlN exceeding 2 ms at helium temperatures and, relevant to spintronic devices, spin lifetimes up to 0.5 $\mu s$ at room temperature.

\end{abstract}
\volumeyear{year}
\volumenumber{number}
\issuenumber{number}
\eid{identifier}
\date[Date text]{date}
\received[Received text]{date}

\revised[Revised text]{date}

\accepted[Accepted text]{date}

\published[Published text]{date}

\startpage{1}
\endpage{102}
\maketitle


Much of semiconductor research in recent years has focused on the electron
spin degree of freedom.
For many non-quantum coherent spintronic
applications one requires \textit{mobile} electrons. \ Even as a
carrier of classical information the spin of mobile electrons offers
advantages: translation and rotation are in principle dissipationless,
offering great potential advantages over charge motion. \ For all types of
spintronic applications, long spin coherence times and ease of spin manipulation 
at \emph{room temperature} are of
crucial importance.
Several types of devices based on mobile spins have been proposed: ballistic
\cite{datta} and non-ballistic\cite{schliemann} spin field effect
transistors, and double-barrier\cite{hall} structures.
Tuning of the spin-orbit (SO) parameters is possible by applying a
gate voltage to a quantum well (QW) to vary the Rashba coupling. \ In
addition, systematic variation of the well width has made it possible to
independently tune other couplings and observe momentum-dependent relaxation
times in {(001)-GaAs QWs.\cite{koralek} 

Experimental studies relevant to the realization of
these devices have mainly been carried out at rather low temperatures $T$ or short
spin lifetimes $\tau_{s}$.
In GaAs, signs of enhanced lifetime (due to the persistent spin 
helix\cite{bernevig}) decreased rapidly 
with temperature and at $T$
= 300 K, $\tau_{s}$ was only slightly above 100 ps; the loss of coherence is due to
cubic-in-k terms in the SO Hamiltonian that relax the spin by the
D'yakonov-Perel' (DP) mechanism.\cite{dyakonov1}
Theory predicts zinc-blende (zb) (111) QWs benefit at $T=0$ K from certain cancelations of SO terms.\cite{cartoixa1}
No experiments on this type of QW have been performed.
There is experimental evidence that (110)-GaAs QWs have enhanced spin coherence times and theory predicts 
infinite spin coherence for one element of the relaxation tensor.\cite{dyakonov2}
However, the strong anisotropy of the spin relaxation tensor ensures rapid decoherence in even 
small magnetic fields.\cite{dohrmann}
At higher temperatures, long spin lifetimes are limited by the 
strength of the cubic-in-k 
SO interaction in direct band gap zb semiconductors. 
In (110) zb-GaAs, spin relaxation times are limited by intersubband spin relaxation at room temperature.\cite{dohrmann}

Nearly all of the theory and experiment along
this research direction has been carried out on materials with the
zb structure. However, there has also been work on
bulk \textit{wurtzite} (w) materials because their SO splittings are small
\cite{beschoten} and there is hope of room temperature ferromagnetism in
magnetically doped w-GaN and w-ZnO.\cite{liu1} \ Spin dynamics in bulk w-GaN
has been studied by two groups,\cite{beschoten} and there has
been experimental \cite{ghosh} and theoretical \cite{harmon, lu} work
on spin lifetimes in w-ZnO. This strongly suggests that w-QWs should be studied,
and indeed there is some recent work along these lines. \ Experimentally,
the SO splittings have been measured by Lo \textit{et al}. \cite{lo} in
Al$_{x}$Ga$_{1-x}$N/GaN QWs by either Shubnikov-de Haas or weak
antilocalization (WAL) measurements. \ \ WAL measurements unambiguously point
to SO coupling \cite{thillosen} and such measurements are found to agree with
theory.\cite{lisesivdin}

Here we determine the 
general temperature dependence of spin lifetimes in w-QWs due to the DP mechanism.  
The DP mechanism is
dominant at room temperature in bulk zb-GaAs\cite{putikka} and w-ZnO\cite{harmon}
and this is expected to be true in modulation doped QWs as well. 
We discuss other relaxation
mechanisms that may take over when DP is suppressed and show that they play no role at room temperature.
For low-Z w-QWs long spin lifetimes of conduction electrons are possible at low and room
temperature due to the combined effects of the w band structure and a small cubic-in-k
coefficient in the SO Hamiltonian.  We also discuss the feasibility of the needed tunings.

There are striking differences in spin relaxation properties between zb symmetry materials and the hexagonal
symmetry of w materials like GaN, ZnO, or AlN. The SO
Hamiltonian $H_{SO}$ for (001) QWs with zb symmetry is
\begin{align}
H_{SO}^{zb} &  =\alpha_{R}(k_{y}\sigma_{x}-k_{x}\sigma_{y})+\beta^{(zb)}_{D}%
(-k_{x}\sigma_{x}+k_{y}\sigma_{y}){}\nonumber\\
&  {}+\beta_{3}(k_{x}k_{y}^{2}\sigma_{x}-k_{y}k_{x}^{2}\sigma_{y}),\nonumber
\end{align}
where $\boldsymbol{\sigma}$ are the Pauli matrices; \ $\alpha_{R}$ is the Rashba
coupling;
$\beta_{3}$ is the cubic-in-k coupling, while $\beta^{(zb)}_{D}=\beta_{3}\left\langle
k_{z}^{2}\right\rangle $ is a Dresselhaus-type term that is controlled by
confinement. $\ \left\langle k_{z}^{2}\right\rangle $ is the expectation value
of the operator $k_{z}$ in the QW wavefunction. \ If $L$ is the QW
width$,$ then $\left\langle k_{z}^{2}\right\rangle \sim\left(  \pi/L\right)
^{2}$ for the lowest electric subband and small structural asymmetry.
$\alpha_{R}$ is proportional to the electric field $E_{z}$ and can thus be
tuned by applying a gate voltage or producing structures with
asymmetry; $\beta^{(zb)}_{D}$ depends on $L$; $\beta_{3}$ depends only on the
material and cannot be turned off. When $\alpha_{R}=\pm\beta^{(zb)}_{D}$ the
linear-in-k terms produce a k-independent effective magnetic field in the
$(110)$ or $(\overline{1}10)$ direction.\cite{schliemann} This can enhance
spin coherence for spins oriented along the effective magnetic field.

For (001) QWs with w symmetry we have\cite{lo, wang}\
\[
H_{SO}^{w}=\left(  \alpha_{R}+\beta^{(w)}_{D}-\beta_{3}k_{\parallel}^{2}\right)
\left(  k_{y}\sigma_{x}-k_{x}\sigma_{y}\right)  ,
\]
where $\beta^{(w)}_{D}=\beta_{1}+b\beta_{3}\left\langle k_{z}^{2}\right\rangle $. 
\ This form is clearly different from the zb case,
as has been confirmed experimentally.\cite{thillosen} As before,
$\alpha_{R}$ can be tuned by applying a gate voltage or varying the
asymmetry; however, $\beta_{1}\neq0$ even in the absence of an applied
electric field - the w structure does not have the mirror symmetry
$z\leftrightarrow-z.$ $\ \beta^{(w)}_{D}$ can be tuned by changing $L$. \ 

There are two important formal differences between $H_{SO}^{zb}$ and
$H_{SO}^{w}.$ \ First, at the linear-in-k level, setting $\alpha_{R}
=-\beta^{(w)}_{D}$ in $H_{SO}^{w}$ gives \textit{zero} effective magnetic field for
all\textit{ }$\boldsymbol{k}$, which means much more dramatic enhancement of spin
coherence. \ As pointed out by Lo et al., this is what makes their Al$_{x}
$Ga$_{1-x}$N/GaN structure an excellent candidate for the non-ballistic spin
field effect transistor.\cite{lo} Second, if we define $\boldsymbol{k}_{\parallel}=\left(  k_{x},k_{y}\right)  $,
then at the cubic-in-k level for a circular Fermi surface
and elastic scattering, $| \boldsymbol{k}_{\parallel}|$ is conserved and we can set
$k_{\parallel}=k_{F},$ the Fermi wavevector.
As pointed out by Wang \textit{et al.}, the effective field can be cancelled \textit{all the way to third order} by enforcing
the condition $\alpha_{R}+\beta^{(w)}_{D}-\beta_{3}k_{F}^{2}=0$, eliminating what
appears to be the major source of spin decoherence in the experiments to date.\cite{lo, wang}
\ Note that the final term can be independently tuned by changing the electron
density. 
The SO Hamiltonian for (111) zb-QWs is similar to (001) w-QWs;\cite{cartoixa1}
similarities also exist between (110) zb and (100)/(010) w-QWs which will be dealt with in a future publication. 

In addition to different crystal symmetries, an important quantitative difference between 
zb and w semiconductors 
is the strength of the SO interaction. The common wurtzite GaN has smaller SO coupling  
than GaAs.  Of prime 
importance for spin lifetimes at room temperature is the coefficient for the cubic-in-k terms
in the SO Hamiltonian; $\beta_3 = -0.32$ meV nm$^3$ for GaN\cite{fu} compared to $\beta_3 = 6.5-30$ meV nm$^3$ for GaAs.\cite{koralek}  From the spin lifetimes evaluated below, an expression for the maximum 
$\tau_s$ at high temperatures, $T\gg T_F$, can be obtained analytically
$\tau^{-1}_{s}(\max)=32\tau_{tr}
m^{\ast 3}\beta_{3}^{2}k_{B}^{3}T^{3}/\hslash^{8}$,
which shows that $\beta_3$ determines
the fall-off of $\tau_{s}(\max)$.
w-AlN is the most favorable, with $\beta_3 = -0.01$ meV nm$^3$, but is less well-characterized than GaN since experiments are lacking and
only one theoretical estimate of $\beta_1$ has been calculated.\cite{nicklas, wang, private}

In Fig. 1(a) we plot $\tau_{s}$ as a function of $\alpha_{R}$ at $T = 300$ K for w-AlN and w-GaN, with the aforementioned parameter values. 
The SO
couplings of w-GaN are the best-characterized of the w materials. \ When
$\alpha_{R}$ is appropriately tuned for w-GaN, $\tau_{s}$ approaches 10 $\mu s$ at $T$=
5 K if $\tau_{tr} = 0.1$ ps and even at $T$ = 300 K can reach values of 4 ns if $\tau_{tr} = 0.1$ ps. 
At $T = 300$ K and $\tau_{tr} = 0.1$ ps,
the maximum spin lifetime in w-AlN is 0.5 $\mu$s. 
At 5 K, the spin lifetime in w-AlN surpasses 2 ms if $\tau_{tr} = 1$ ps. 
The peaks in $\tau_{s}$ lie
slightly off the condition $\alpha_{R}=-\beta^{(w)}_{D}$; the
difference is a measure of the importance of the cubic-in-k term. 
This
implies that devices meant to be operated at different temperatures would need
to be tuned somewhat differently. 
The extremely long spin relaxation times in both w-GaN and w-AlN are not limited by the Elliott-Yafet (EY) 
mechanism\cite{elliott} since we determine the spin relaxation time due to EY to be $\tau_{EY} \sim 100 \mu$s at room temperature. 
Unlike in (110) zb-QWs, intersubband spin relaxation is not the limiting mechanism either; due to the small SO coupling we find it to 
be  1 ms - $10^6$ times weaker than in zb-GaAs QWs.\cite{dohrmann}
As a comparison to the w materials we show the corresponding high temperature calculations for (001) and (111) zb-GaAs QWs in Fig. 1(b). The
times are shorter by orders of magnitude compared to the w materials.
High temperature maxima expressions for zb-(001) and zb-(111) are similar to what was determined for w-(001). The
contrast in $\tau_s$ is due to the much larger $\beta_3$ in GaAs. 

We now sketch the calculation that gives the results quoted above.
The spin relaxation tensor $\Gamma_{ij}$ is defined by the equation
$dS_{i}/dt=-\sum_{j=x,y,z}\Gamma_{ij}S_{j}$ and $\boldsymbol{S}$ is the spin
polarization vector. \ The DP contribution to $\Gamma$ is given by
\cite{kainz}
\begin{equation}
\Gamma_{ij}(T)  =\frac{1}{2\hbar^{2}}\sum_{n=-\infty}^{\infty
}\frac{\int_{0}^{\infty}dE[f_{+}(  E,T)  -f_{-}(
E,T)]  \tau_{n}\gamma_{n}^{ij}}{\int_{0}^{\infty}dE[
f_{+}(  E,T)  -f_{-}(  E,T)]},\nonumber
\end{equation}
where $\gamma_{n}^{ij}(  \boldsymbol{k}_{\parallel})  $= Tr$\{
[  H_{SO}^{(-n)},[ H_{SO}^{(n)
},\sigma_{j}]]  \sigma_{i}\}$. The scattering rates are given by $1/\tau_{n}$=$\int_{0}^{2\pi}
d\theta ~W(  E,\theta) (  1-\cos n\theta)$. $H_{SO}^{(
n)  }$=$\int_{0}^{2\pi}d\phi_{k}H_{SO}(\boldsymbol{k}_{\parallel})
\exp(  -in\phi_{k})$ are the harmonics of the SO Hamiltonian. Here $f_{\pm}(E,T)  $ is the Fermi distribution for spins of positive
and negative helicity $\left\langle \boldsymbol{\sigma}\cdot\boldsymbol{k}_{\parallel
}\right\rangle ,$ $\phi_{k}$ is the angle between $\boldsymbol{k}_{\parallel}$ and
the $k_{x}$-axis, $\theta$ is the scattering angle, and $W\left(
E,\theta\right)  $ is the scattering rate. \ The various
scattering times are energy-dependent due to the energy dependence of $W.$
Here we take $\tau_1$ to be energy independent for
simplicity and to illustrate the main ideas. We reserve a full treatment
of the energy dependence for future work.

We focus first on the w case with a circular Fermi surface.
When the energy splitting between positive and negative helicity states is
small, we find
\begin{align}
\Gamma_{x,y}^{(w)}(T) & = 
\frac{1}{2}\Gamma_{z}^{(w)}(T) = 
\frac{2\tau_{tr}}{\hslash^{2} I_{0}(\beta\mu) }
\big[(\alpha_{R}+\beta^{(w)}_{D})^{2} \zeta I_{1}(\beta\mu){}\nonumber\\
& {} - 2(\alpha_{R}+\beta^{(w)}_{D})
\beta_{3}\zeta^2
I_{2}(\beta\mu) +\beta_{3}^{2}\zeta^3 I_{3}(\beta\mu)\big]\label{eq:gw},
\end{align}
where $\zeta=2m^{\ast}k_{B}T/\hbar^{2}$, $\beta=1/k_{B}T,$ $\mu$ is the chemical potential, $I_{r}\left(
z\right)  \equiv\int_{0}^{\infty}dx~x^{r}/\left\{  4\cosh^{2}\left[  \left(
x-z\right)  /2\right]  \right\} $, and $\tau_{tr}$, the transport time,  is an experimental quantity.  
All other components of $\Gamma$ vanish.
This simplifies at zero temperature
$\Gamma_{x,y}^{\left(  w\right)  } = 4\tau_{1}m^{\ast
}E_{F}(  \alpha_{R}+\beta^{(w)}_{D}-2m^{\ast}\beta_{3}E_{F}\hbar^{-2})  ^{2}\hbar^{-4}$.
$E_{F}$ is the Fermi energy. \ Clearly the $T=0$ relaxation times diverge when
the tunable quantity $\alpha_{R}+\beta^{(w)}_{D}-2m^{\ast}\beta_{3}E_{F}/\hbar^{2}$
vanishes. \ This divergence is cut off by finite temperatures. 

In zb structures $\Gamma$ is observed to be anisotropic,
and the appropriate quantities are $\Gamma_{+\left(  -\right)  }^{(zb)}$ and $\Gamma_z^{(zb)}$ the
relaxation rates for spin along the [110] ([$\overline{1}$10]) and [001] directions.
The results are determined in Ref. \onlinecite{kainz}.
These
expressions for zb approach finite limits as $T\rightarrow0,$ \textit{regardless of
the values of the parameters}. \ This is in sharp contrast to the relaxation
rates in Eq. \ref{eq:gw}.

We now address the tunability of possible devices. In
zb-GaAs it has been possible to achieve quite substantial variations in the
appropriate parameters.\cite{koralek, studer} Koralek et al.
were able to change $\beta^{(zb)}_{D}$ by making structures with different well
widths and to change $\alpha_{R}$ by adjusting the dopant concentrations on
the sides of the well, corresponding to a maximum electric field of $5.4 \times
10^{-3}~$V$/$nm. \ In this way a range of $\alpha_{R}/\beta^{(zb)}_{D}$ of about
0.25 to 1.25 was achieved, without even needing a gate. \ $\beta_{3}$ was
inaccessible and remained constant for all structures, setting an upper limit
on spin lifetimes. \ Significant experimental tuning of SO coupling in w-GaN
has not yet been achieved. However, calculations have been done\cite{lo2}
for w-GaN, which produce the correct magnitude for the spin splittings overall
($\sim5$ meV at Fermi wavectors of typical structures). These authors do not
compute $\alpha_{R},$ $\beta_{1},$ and $\beta_{3}$ explicitly, but their
computed spin splittings at a typical Fermi wavevector shows that changes in
spin splittings by a factor of 4 or so can be achieved by changing the well
width from 10 to 2 unit cells; to achieve the same sort of change due to
external electric fields required very strong fields of order 1 V$/$nm. This
suggests that changing well width and electron density rather than electric
field will be the most favorable route for tuning of w-QWs.

In conclusion, we predict that QWs consisting of w materials, e.g. GaN/GaAlN,
can be tuned to achieve very long spin lifetimes. \ These lifetimes are at
zero momentum ($\boldsymbol{q}=0)$, not helical modes, and are therefore better for
spin injection and transistor devices operating at mesoscopic length scales.
We predict AlN may offer the longest spin lifetimes, with a
lifetime of 0.5 $\mu$s at room temperature.

We would like to thank E. Johnston-Halperin, J. Nicklas, and Meng-En Lee for
useful conversations. \ Financial support was provided by the National Science
Foundation, Grant Nos. NSF-ECS-0524253 (RJ), NSF-FRG-0805045 (RJ), and NSF-ECS-0523918 (NH and WP).

\begin{figure}[ptbh]
\label{fig:gan}  \begin{centering}
        \includegraphics[scale = 0.35,trim = 10 5 30 00, angle = 0,clip]{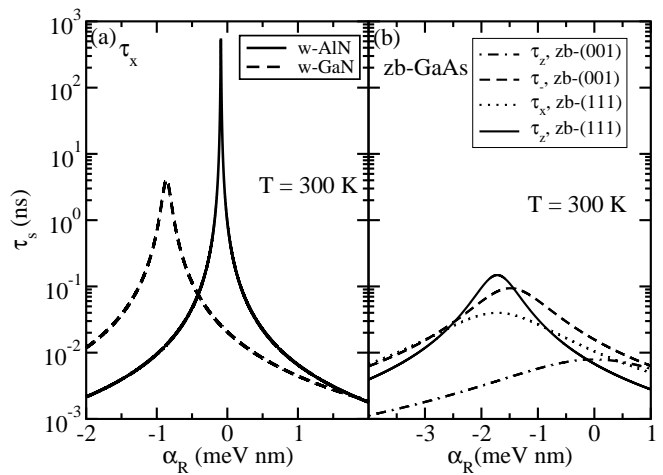}
        \caption[]
{(a) Spin relaxation times $\tau_{y} = \tau_{x}$ (Eq. \ref{eq:gw}) for w-AlN and w-GaN;
(b) spin relaxation times for various zb-GaAs QW structures for zb-GaAs as a function of Rashba
coupling $\alpha_{R}$ at $T = 300$ K.
Parameters used for w-AlN: $\beta_{1} = 0.09$ meV nm,\cite{nicklas} $T_{F} = 15$ K;
w-GaN: $\beta_{1} = 0.9$ meV nm,\cite{fu} $T_{F} = 25$ K;
zb-GaAs: $\beta_{3} = 20$ meV nm$^{3}$, $T_{F} = 80$ K. 
$L = 10$ nm and $\tau_{tr} = 0.1$ ps for all QWs. 
All $T_F$'s correspond to an electron density $\sim 2 \times10^{11}$ cm$^{-2}$.}
        \end{centering}
\end{figure}


\ \ \ \ \ \ \ \ \ \ \ \

\end{document}